\documentclass[12pt,nofootinbib]{revtex4-1}

\usepackage[utf8]{inputenc}
\usepackage{amsmath}
\usepackage{amssymb}
\usepackage{amsfonts}
\usepackage{graphicx}
\usepackage{color}
\usepackage{amsbsy}
\usepackage{url}
\usepackage{enumerate}
\usepackage[vcentermath,enableskew]{youngtab}
\usepackage{verbatim}
\usepackage[normalem]{ulem}
\usepackage{cancel}
\usepackage{bm,bbm}
\usepackage{slashed}

\usepackage{setspace}

\makeatletter
\@addtoreset{equation}{section}

\makeatother

\newcommand{\sech}{\mbox{sech}}

\def\be{\begin{eqnarray}}
\def\ee{\end{eqnarray}}
\def\p{\partial}

\begin{document}
\vspace*{-1cm}
\begin{flushright}
YGHP-18-03
\end{flushright}

\title{
Localization of the Standard Model via Higgs mechanism and a 
finite electroweak monopole
from non-compact five dimensions
}

\author{Masato Arai$^1$, Filip Blaschke$^{2, 3}$, Minoru Eto$^4$ 
and Norisuke Sakai$^5$\ \\\ }
\affiliation{
$^1$Faculty of Science, Yamagata University, 
Kojirakawa-machi 1-4-12, Yamagata,
Yamagata 990-8560, Japan\\
$^2$Faculty of Philosophy and Science, Silesian University in 
Opava, Bezru\v{c}ovo n\'am. 1150/13, 746~01 Opava, Czech Republic\\
$^3$Institute of Experimental and Applied Physics, Czech Technical 
University in Prague, Horsk\'a 3a/22, 128 00 Praha 2, Czech Republic\\
$^4$Department of Physics, Yamagata University, 
Kojirakawa-machi 1-4-12, Yamagata,
Yamagata 990-8560, Japan\\
$^5$Department of Physics, and Research and 
Education Center for Natural Sciences, 
Keio University, 4-1-1 Hiyoshi, Yokohama, Kanagawa 223-8521, Japan\\
and iTHEMS, RIKEN,
2-1 Hirasawa, Wako, Saitama 351-0198, Japan
}

\begin{abstract}
\ \\
We propose a minimal and self-contained model in non-compact 
flat five dimensions which localizes the Standard Model (SM) 
on a domain wall. 
Localization of gauge fields is achieved by the condensation 
of Higgs field via a Higgs dependent gauge 
kinetic term in five-dimensional Lagrangian.
The domain wall connecting vacua with unbroken gauge 
symmetry drives the Higgs condensation which provides both 
electroweak symmetry breaking and gauge field localization 
at the same time. 
Our model predicts higher-dimensional interactions 
$|H|^{2n}(F_{\mu\nu})^2$ in the low-energy effective theory.
This leads to two expectations: The one is a new tree-level 
contribution to $H \to \gamma\gamma$ ($H \to gg$) decay whose signature 
is testable in future LHC experiment. 
The other is a finite electroweak monopole which may be
accessible to the MoEDAL experiment. 
Interactions of translational Nambu-Goldstone boson is 
shown to satisfy a low-energy theorem. 

\end{abstract}

\maketitle

\newpage

%


\section{Introduction}

The hypothesis that our four-dimensional world is embedded in 
higher-dimensional spacetime has been a hot topic in high energy 
physics for decades.
Indeed, many mysteries of the Standard Model (SM) can be explained 
in this way.
In particular, the discovery of D-branes in superstring theories 
\cite{Polchinski:1995mt} has intensified the research of the 
brane-world scenarios more than anything else. 
Then the seminal works \cite{ArkaniHamed:1998rs,Antoniadis:1998ig,
Randall:1999ee,Randall:1999vf} provided the basic templates 
for further studies.

The biggest advantage of models in extra dimensions is to utilize 
{\it geometry} of the extra dimensions.
A conventional setup, common among the extra-dimensional models, 
is that extra dimensions are prepared as a compact 
manifold/orbifold. Namely, our four-dimensional spacetime is 
treated differently compared with extra dimensions. 

In order to make things more natural, we can harness the {\it topology} 
of extra dimensions in addition to the geometry.
The idea is quite simple and dates back to early 80's 
\cite{Rubakov:1983bb}, namely that the seed of dynamical 
creation of branes in extra-dimensions is a spontaneous 
symmetry breaking giving rise to a topologically stable soliton/defect 
on which our four-dimensional world is localized. 
The topology ensures not only stability of the brane but also 
the presence of chiral matters localized on the brane 
\cite{Jackiw:1975fn,Rubakov:1983bb}.
In addition, graviton can be trapped \cite{Cvetic:1992bf,DeWolfe:1999cp,
Csaki:2000fc,Eto:2002ns,Eto:2003bn,Eto:2003ut}.
Thus, the topological solitons provide a natural framework 
bridging gap between extra dimensions and four dimensions.

In contrast, localizing massless gauge bosons, especially non-Abelian 
gauge bosons, is quite difficult. There were many works so far 
\cite{Dvali:2000rx, Kehagias:2000au, Dubovsky:2001pe, 
Ghoroku:2001zu,Akhmedov:2001ny, Kogan:2001wp, Abe:2002rj, 
Laine:2002rh, Maru:2003mx, Batell:2006dp, Guerrero:2009ac, 
Cruz:2010zz, Chumbes:2011zt, Germani:2011cv, Delsate:2011aa, 
Cruz:2012kd, Herrera-Aguilar:2014oua, Zhao:2014gka, Vaquera-Araujo:2014tia,
Alencar:2014moa,Alencar:2015awa,Alencar:2015oka,Alencar:2015rtc,Alencar:2017dqb}. 
However, each of these has some advantages/disadvantages and 
there seems to be only little universal understanding. 
Then, a new mechanism utilizing a field dependent gauge kinetic 
term (field dependent permeability)  
\be
- \beta(\phi_i)^2 F_{MN}F^{MN}, \quad (M,N=0,1,2,3,4),
\label{eq:f2}
\ee
came out in Ref.~\cite{Ohta:2010fu} where $\phi_i$ are scalar 
fields. 
This is a semi-classical realization of the confining phase  
\cite{ArkaniHamed:1998rs,Dvali:1996xe,Kogut:1974sn,Fukuda:1977wj, 
Luty:2002hj, Fukuda:2009zz, Fukuda:2008mz} rather than Higgs 
phase outside the solitons.
The authors have continuously studied brane-world models with 
topological solitons by using (\ref{eq:f2}) \cite{Arai:2012cx,
Arai:2013mwa,Arai:2014hda,Arai:2016jij,Arai:2017lfv,Arai:2017ntb,
Arai:2018rwf}.
Let us highlight several results: We investigated the geometric 
Higgs mechanism which is the conventional Higgs mechanism driven 
by the positions of multiple domain walls in an extra dimension 
in Ref.~\cite{Arai:2017ntb}.
Then we proposed a model in which the brane world on five domain 
walls naturally gives $SU(5)$ Grand Unified Theory  
in Ref.~\cite{Arai:2017lfv}. Furthermore, we have clarified how 
to derive a low-energy effective theory on the solitons 
in the models with a non-trivial gauge kinetic term (\ref{eq:f2}) 
by extending the $R_\xi$ gauge in any spacetime dimensions 
$D$ \cite{Arai:2018rwf}.
Another group also recently studied the SM in a similar model 
with $\beta^2$ taken as a given background in $D=5$ 
\cite{Okada:2017omx,Okada:2018von}. They have also discussed 
phenomenology involving Nambu-Goldstone (NG) bosons for broken 
translation.

In this paper, we propose a minimal and self-contained model 
in non-compact flat five dimensions which localizes the SM 
on a domain wall.
A striking difference from the previous works \cite{Arai:2012cx,
Arai:2013mwa,Arai:2014hda,Arai:2016jij,Arai:2017lfv,Arai:2017ntb,
Arai:2018rwf} is that we do not 
need extra scalar fields $\phi_i$ which were introduced only 
for localizing gauge fields via Eq.~(\ref{eq:f2}). 
Instead, we put the SM Higgs in that role. As a consequence, 
localization of massless/massive gauge fields and the electroweak 
symmetry breaking have the same origin. In other words, the 
Higgs field is an active player in five dimensions with a new 
role as a localizing agent of gauge fields on the domain wall, 
in addition to the conventional roles giving masses to gauge 
bosons and fermions. 
Since our model does not need extra scalar fields $\phi_i$, 
it is not only very economical in terms of field content but also
we are free from a possible concern that 
$\phi_i$ would give an undesirable impact on the
low-energy physics. 
We also study the translational NG 
boson $Y(x^\mu)$. Due to a low-energy theorem, it should have 
a derivative coupling with all other particles 
including Kaluza-Klein (KK) particles. We find a new vertex 
$\bar \psi^{(KK)}\gamma^\mu\p_\mu Y\psi^{(SM)}$ which
provides a new diagram for 
the production of KK quarks  
$\psi^{(SM)}+\psi^{(SM)}\to \psi^{(KK)}+\psi^{(KK)}$ in the LHC 
experiment. 
This should be a dominant production process compared to the usual gluon
fusion, and can easily violate experimental bounds. 
To avoid this, we will set a fundamental five-dimensional energy scale sufficiently large, providing all the KK modes supermassive.
However, surprisingly, the Higgs dependent gauge kinetic term (\ref{eq:f2}) can naturally leave masses 
of  localized lightest particles to be of order the SM energy scale.
Thus, all KK particles and the NG boson have no impact on the 
low-energy physics.
Nevertheless, 
as a consequence of Eq.~(\ref{eq:f2}), regardless of the extra particles, our model still has a new 
experimental signature in $H \to \gamma\gamma$ ($H\to gg$) decay channel 
at tree level
which is testable in future LHC experiment.
Furthermore, we point 
out that the localization via Eq.~(\ref{eq:f2}) yields higher 
dimensional interactions $|H|^{2n}(F_{\mu\nu})^2$ in the low-energy 
effective theory and it  provides a natural reason to have a 
finite electroweak monopole solution. 
Its mass has been previously estimated \cite{Cho:2013vba,Ellis:2016glu} as $\lesssim 5.5$ TeV, 
so that it could be pair-produced at the LHC and accessible to 
the MoEDAL experiment \cite{Acharya:2016ukt,Acharya:2017cio}.
Thus, our model can pay the price for an electroweak monopole.

The paper is organized as follows. 
In Sec.~\ref{sec:2}, we explain all the essential ingredients 
in a simple toy model of Abelian-Higgs-scalar model in $D=5$.
We explain how a domain wall drives condensation of the Higgs 
field and at the same time localizes massless/massive gauge bosons 
and also chiral fermions. Phenomenological viability, the translational zero mode, and relevance 
of the  $H\to \gamma\gamma$ decay channel are addressed in 
Sec.~\ref{sec:3}. 
We present a realistic model localizing the SM in Sec.~\ref{sec:4} 
and discuss the finite electroweak monopole in Sec.~\ref{sec:5}.
Our results are summarized and discussed in Sec.~\ref{sec:6}. 
Appendix \ref{app:mode_eq} is devoted to define mode 
expansion on the stable background. Mode expansion and effective 
potential on unstable background is given in 
Appendix \ref{sc:modeEq_unstable_sol}.
Some formulae  for KK fermion pair production by NG boson exchange
are given in Appendix \ref{sc:cross}.

\section{Localization via Higgs mechanism}
\label{sec:2}

In order to illustrate a novel role of the Higgs mechanism 
besides the conventional roles of giving masses to gauge fields 
and chiral fermions in a gauge invariant manner, let us consider 
a simple Abelian-Higgs-scalar model in $D=5$ flat spacetime 
as a toy model. 
The following arguments are quite universal so that it is 
straightforward to apply them to non-Abelian gauge theories, 
such as the SM which we discuss in Sec.~\ref{sec:4} and 
also to models with $D\ge5$ \cite{Arai:2018xyz}.

A simple Abelian-Higgs-scalar model in $D=5$ reads:\footnote{
The bosonic part is a simple extension of the well-studied model 
\cite{Montonen,Sarkar,Ito} in which the Higgs field ${\cal H}$ is 
replaced by a real scalar field.}
\be
{\cal L} &=& - \beta({\cal H})^2 {\cal F}_{MN}^2 
+ |{\cal D}_M {\cal H}|^2 + (\p_M{\cal T})^2 
- V \nonumber\\
&& + i \bar\Psi\Gamma_M {\cal D}^M\Psi + i \bar{\tilde\Psi}\Gamma_M \p^M\tilde\Psi + \left(\eta {\cal T} \bar\Psi\Psi
- \tilde\eta {\cal T} \bar{\tilde\Psi}\tilde \Psi + \chi{\cal H}\bar \Psi \tilde\Psi +{\rm h.c.}\right), 
\label{eq:Lag}\\
V &=& \Omega^2 |{\cal H}|^2 + \lambda^2\left(|{\cal H}|^2 
+ {\cal T}^2 - v^2\right)^2,
\label{eq:pot5}
\ee
with ${\cal F}_{MN} = \p_M {\cal A}_N - \p_N {\cal A}_M$. 
Here ${\cal T}$ is a real scalar field, and 
${\cal H}$ is the Higgs field which interacts with ${\cal A}_M$ 
not only via the covariant derivative 
${\cal D}_M {\cal H}= \p_M {\cal H}  + i q_H{\cal A}_M {\cal H}$,
but also through non-minimal gauge kinetic term with the 
field-dependent function $\beta^2$ defined by
\be
\beta({\cal H})^2 = \frac{|{\cal H}|^2}{4\mu^2}.
\label{eq:beta_simple}
\ee
The covariant derivative of the charged fermion field is defined 
by ${\cal D}_M \Psi = \p_M \Psi + i q_f {\cal A}_M \Psi$.
$\tilde \Psi$ is a neutral fermion.
The bosonic part of the model has $Z_2$ symmetry 
${\cal T} \to - {\cal T}$. 
Mass dimensions of the fields and parameters are summarized as 
$[{\cal H}]=[{\cal T}]=\frac{3}{2}$, $[{\cal A}_M]=1$, $[\Psi]= [\tilde\Psi]= 2$, 
$[\mu] = [\Omega] = [\lambda^{-2}] = [\eta^{-2}] = [\tilde \eta^{-2}] = [\chi^{-2}] = [v^\frac{2}{3}] = 1$, 
and $[\beta]=\frac{1}{2}$.
The five-dimensional Gamma matrix $\Gamma^M$ is related to four-dimensional one as
$\Gamma^\mu = \gamma^\mu$ and $\Gamma^5 = i\gamma^0\gamma^1\gamma^2\gamma^3= i\gamma^5$.

There are two discrete vacua ${\cal T} = \pm v$ with ${\cal H}=0$. 
The vacua break the $Z_2$ symmetry but preserve $U(1)$ gauge 
symmetry which is necessary to localize the massless $U(1)$ 
gauge field on a domain wall \cite{Ohta:2010fu,Arai:2012cx,
Arai:2013mwa,Arai:2014hda,Arai:2016jij,Arai:2017lfv,Arai:2017ntb,
Arai:2018rwf}. 
Therefore, the Higgs mechanism does not take 
place in the vacua.

However, spontaneous breaking of the $Z_2$ symmetry gives 
rise to a topologically stable domain wall, connecting these 
two discrete vacua. 
Depending on the values of the parameters, 
the following stable domain wall solutions are obtained 
\be
{\cal T}_0{}' &=& v \tanh \lambda v y,\quad {\cal H}_0 = 0,
\qquad (\lambda v \le \Omega),
\label{eq:T0}\\
{\cal T}_0 &=& v \tanh \Omega y,\quad {\cal H}_0 
= \bar v\, \sech\,\Omega y ,\qquad (\lambda v > \Omega),
\label{eq:T1}
\ee
with $\bar v = \sqrt{v^2 - \Omega^2/\lambda^2}$ and $y = x^4$.
We are not interested in the former solution (\ref{eq:T0}) 
since the $U(1)$ is unbroken everywhere and the gauge field 
is not dynamical due to $\beta^2 = 0$.
On the other hand, as we will show below, the latter solution 
(\ref{eq:T1}) localizes the $U(1)$ gauge field by 
$\beta^2 \propto\,\sech^2\Omega y$. 
When the Higgs is neutral ($q_H =0$), the lightest mode of 
the localized gauge field is precisely massless \cite{Arai:2017lfv,
Arai:2017ntb,Arai:2018rwf} whereas, as we will see, it becomes 
massive when the Higgs is charged ($q_H \neq 0$).

To understand the mechanism for the localized massless 
gauge field to become massive, let us compute the low-energy 
effective potential for the effective Higgs field in four dimensions 
in the parameter region 
\begin{equation}
0<\epsilon^2 \ll 1,\quad 
\epsilon^2 \equiv \frac{\lambda^2 \bar v^2}{\Omega^2}
=\frac{\lambda^2 v^2-\Omega^2}{\Omega^2}. 
\label{eq:parameter_region}
\end{equation}
From the linearized field equation around 
the background of the domain wall solution (\ref{eq:T1}), 
we find that there is a mass gap of order $\Omega$, and two 
discrete modes much lighter than the mass gap. 
The lowest mode is exactly massless Nambu-Goldstone (NG) boson 
corresponding to spontaneously broken translation symmetry 
along the $y$ direction. Its interactions with all other 
effective fields are generally suppressed by inverse powers 
of large mass scale, whose characteristics will be discussed in 
Sec.~\ref{eq:translational_zero_mode} and Sec.~\ref{sec:4}. 
Disregarding the NG boson, we retain only one light boson, 
whose wave function is well-approximated by the same functional 
form as the background solution ${\cal H}_0(y)$ in (\ref{eq:T1}). 
When $\lambda \bar v=0$, this wave function gives the zero 
mode exactly, corresponding to the condensation mode at the 
critical point $\lambda v = \Omega$, where the ${\cal H}$ 
field begins to condense. 
After ${\cal H}$ condenses, this mode becomes slightly massive 
above the critical point (\ref{eq:parameter_region}) with the 
mass of order $\lambda \bar v$, whose wave function 
receives small corrections suppressed by powers of $\epsilon$ 
(including an admixture of fluctuations of ${\cal T}$). 
Combining the background solution and the fluctuation, we 
introduce the following effective field $H(x)$ (a quasi-moduli) corresponding 
to the Higgs field in the low-energy effective field theory 
\be
{\cal H}(x, y) = \sqrt{\frac{\Omega}{2}}\,H(x)\, \sech\,\Omega y.
\label{eq:H}
\ee
Inserting this Ansatz into the Lagrangian and integrating over $y$, 
we obtain effective action as 
\be
{\cal L}_{\rm Higgs}(H)
= |D_\mu H|^2  
- V_H, 
\quad
V_H =  \lambda_2^2 |H|^2 + \frac{\lambda_4^2}{2} |H|^4,
\label{eq:L_higgs}
\ee
\be
\lambda_2^2 = - \frac{4\lambda^2 \bar v^2}{3},\quad
\lambda_4^2 = \frac{2\lambda^2\Omega}{3},
\label{eq:constant_higgs}
\ee
where the effective gauge field in the covariant derivative 
$D_\mu$ is more precisely defined below, see 
Eq.~(\ref{eq:gauge_kin_term}). 
The possible corrections suppressed by powers of $\epsilon^2$ 
can be systematically computed as described in 
Appendix~\ref{app:mode_eq}. 
This is just a conventional Higgs Lagrangian which catches all 
the essential features. 
First, note that the sign of the quadratic term is determined by 
$\bar v^2 = v^2 - \Omega^2/\lambda^2$.
When $\bar v^2 = 0$ ($v\lambda = \Omega$), the Higgs is massless 
corresponding to the condensation zero mode in (\ref{eq:H}).
When $\bar v^2 < 0$ ($v\lambda < \Omega$), the vacuum expectation 
value (VEV) is $\left<H\right> =0$. Thus, we reproduce the 
solution (\ref{eq:T0}).
On the other hand, when  $\bar v^2 > 0$ ($v\lambda > \Omega$), 
we have non zero VEV for the effective Higgs field $H(x)$ 
\be
\left<H\right> = \sqrt{\frac{2}{\Omega}}\,\bar v \equiv 
\frac{v_h}{\sqrt2},
\label{eq:vev}
\ee
which correctly gives the solution (\ref{eq:T1}). 
Note that the VEV $v_h$ can also be obtained directly from the 
five-dimensional field ${\cal H}$ as 
\be
\frac{v_h^2}{2} = \int^\infty_{-\infty}dy\, 
{\cal H}_0^2 = \frac{2\bar v^2}{\Omega}.
\ee
The mass of physical Higgs boson can be read from 
Eq.~(\ref{eq:L_higgs}) as 
\be
m_h^2  = \frac{8}{3}\lambda^2 \bar v^2 
= \frac{8}{3}\Omega^2 \epsilon^2,
\label{eq:mh}
\ee
which is of order $\epsilon^2$ as we expected. 
Thus, the $y$-dependent Higgs condensation ${\cal H}_0(y)$ of 
Eq.~(\ref{eq:T1}) in $D=5$ which is driven by the domain wall 
${\cal T}_0(y)$ connecting two unbroken vacua gives indeed the 
Higgs mechanism through Eq.~(\ref{eq:L_higgs}). 
To complete the picture, we next calculate the mass of gauge 
bosons. 
We will assume $v\lambda > \Omega$ in the rest of paper, so 
that the solution (\ref{eq:T1}) always applies.

\begin{figure}[t]
\begin{center}
\includegraphics[width=16cm]{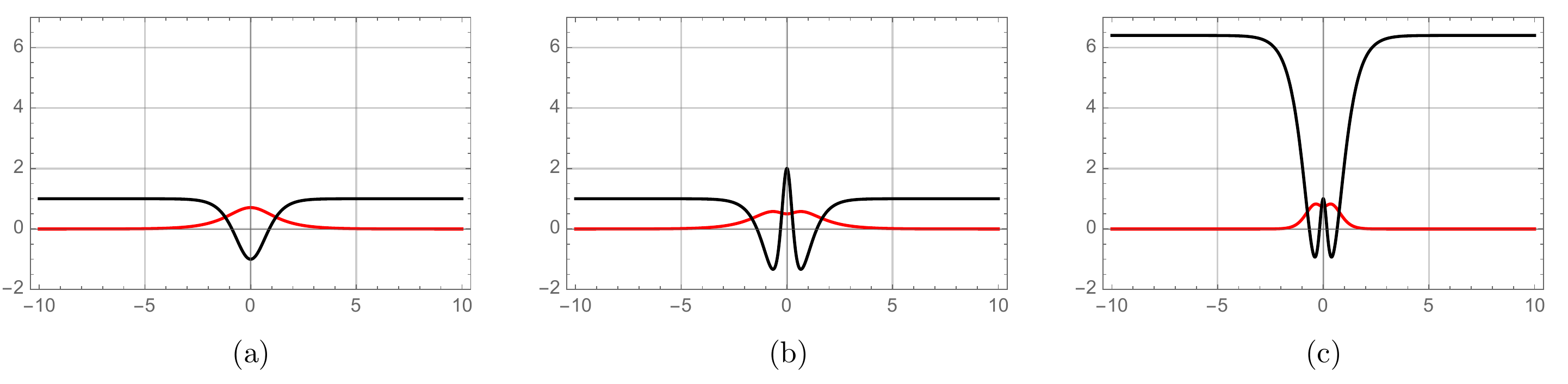}
\caption{The black lines show the Schr\"odinger potentials 
$V_{\rm S} = \beta''/\beta$ for $\beta^2$ given in 
Eq.~(\ref{eq:beta_simple}) (a), 
in Eq.~(\ref{eq:beta_mod}) (b), and  in Eq.~(\ref{eq:beta2_EMY}) (c).
The potential of (c) is multiplied by $0.1$ for clarity. 
The horizontal axis is $\Omega y$.
The red curves show the corresponding zero mode wave functions.}
\label{fig:fig}
\end{center}
\end{figure}

To figure out the spectrum of the gauge field, first of all, 
we use canonical normalization $A_M = 2\beta {\cal A}_M$. 
The linearized equation of motion for $A_\mu$ in the generalized 
$R_\xi$ gauge \cite{Arai:2018rwf,Arai:2018xyz} is 
\be
\left\{\eta^{\mu\nu}\square - \left(1-\frac{1}{\xi}\right)\p^\mu\p^\nu 
+ \eta^{\mu\nu}\left(-\p_y^2 + 
\frac{(\p_y^2\beta)}{\beta} + 2q_H^2 \mu^2 \right)\right\}A_\nu = 0.
\ee
Thus, the Kaluza-Klein (KK) spectrum is identical to eigenvalues 
of 1D quantum mechanical problem with the Schr\"odinger potential 
$V_{\rm S} = (\p_y^2 \beta)/\beta + 2q_H^2\mu^2$. 
Fig.~\ref{fig:fig} (a) shows the corresponding Schr\"odinger potential.
The eigenvalues $m_n^2$ and eigenfunctions $\phi_n(y)$
can be easily obtained \cite{Arai:2018rwf}. 
There is a unique bound state 
\be
\phi_0(y) = \frac{\sqrt{2}\bar v}{v_h}\,\sech\,\Omega y,
\quad m_0^2 = 2q_H^2\mu^2.
\label{eq:phi_0}
\ee
No other bound states exist and a continuum of scattering modes 
parametrized by the momentum $k$ corresponds to the eigenvalues 
$m_k^2 = k^2 + \Omega^2+q_H^2 \mu^2$.
Thus, the mass gap between the unique bound state $\phi_0$ 
and the higher KK modes is of order $\Omega$ (under the assumption 
$\Omega \gg \mu$) which is the inverse width of the domain wall.
In terms of the original field ${\cal A}_\mu$, the lightest 
massive gauge boson $A_\mu^{(0)}(x)$ is given by 
\be
{\cal A}_\mu = \frac{A_\mu}{2\beta} 
= \frac{\mu \phi_0}{{\cal H}_0}A_\mu^{(0)}(x) + \cdots 
= \frac{\sqrt{2}\mu}{v_h} A_\mu^{(0)}(x) + \cdots,
\label{eq:expand_A}
\ee
where the ellipses stand for the heavy continuum modes. 
The mass of the lightest massive gauge boson is 
\be
m_A = m_0 = \sqrt{2}\,q_H\mu.
\label{eq:m_A}
\ee
One can show that the fifth gauge field ${\cal A}_y$ has no 
physical degrees of freedom \cite{Arai:2018rwf}.

Having Eq.~(\ref{eq:expand_A}) at hand, we are now able to 
read the effective gauge coupling constant.
By plugging Eqs.~(\ref{eq:H}) and (\ref{eq:expand_A}) into 
Eq.~(\ref{eq:Lag}) and integrating it over $y$, we have the 
kinetic term for $H$ as 
\be
\int^\infty_{-\infty} dy\ |{\cal D}_\mu {\cal H}|^2 
= \left|\left(\p_\mu + i q_H \frac{\sqrt{2}\mu}{v_h} 
A^{(0)}_\mu\right)H\right|^2
+ \cdots
=|D_\mu H|^2+\cdots,
\label{eq:gauge_kin_term}
\ee
where the ellipses stand for the massive modes. 
Thus, the effective four-dimensional gauge coupling  reads 
\be
e = \frac{\sqrt{2}\mu}{v_h}.
\label{eq:e}
\ee
Combining Eq.~(\ref{eq:e}) with Eq.~(\ref{eq:m_A}), and 
Eq.~(\ref{eq:vev}) with Eq.~(\ref{eq:mh}), we see that what 
happens here is perfectly consistent with the ordinary Higgs 
mechanism 
\be
m_A  
=  q_He v_h,\quad
m_h = \lambda_4 v_h.
\ee

Finally, let us investigate the domain wall fermions $\Psi$ and $\tilde\Psi$
\cite{Jackiw:1975fn,Rubakov:1983bb}. 
In the region (\ref{eq:parameter_region}) together with a phenomenological condition 
explained later, our parameters should satisfy the following inequality
\be
\frac{\eta v}{\Omega} \simeq \frac{\tilde \eta v}{\Omega} \simeq 1 \gg \frac{\chi \bar v}{\Omega}.
\label{eq:yukawa}
\ee
Then we can treat the Yukawa term ${\cal H} \bar\Psi\tilde\Psi$ in Eq.~(\ref{eq:Lag}) as a perturbation.
In order to study the unperturbed Dirac equation, we
decompose five-dimensional fermions as 
\be
\Psi = \sum_n \left(f_L^{(n)}(y) \psi_L^{(n)}(x) 
+ f_R^{(n)}(y) \psi_R^{(n)}(x)\right),
\ee
where $\psi_L^{(n)}$ and $\psi_R^{(n)}$ are left-handed $(\gamma^5\psi_L = - \psi_L$) and 
right-handed $(\gamma^5\psi_R = \psi_R$) spinors in four-dimensions 
\be
i\p\!\!\!/ \psi_L^{(n)} = M_n \psi_R^{(n)},\quad
i\p\!\!\!/ \psi_R^{(n)} = M_n \psi_L^{(n)},
\ee
and the mode functions $f_L^{(n)}$ and $f_R^{(n)}$ satisfies 
\be
Q  f_L^{(n)} + M_n f_R^{(n)} = 0,\quad
Q^\dagger  f_R^{(n)}  +  M_n f_L^{(n)} = 0,
\ee
with $Q = \p_y + \eta {\cal T}_0$, and  $Q^\dagger = -\p_y + \eta {\cal T}_0$.
Assuming the five-dimensional Yukawa coupling to satisfy 
$\eta > 0$, we find a unique zero mode 
\be \label{eq:fermzeromode}
f_L^{(0)}(y) = N_{L,0} \left(\cosh \Omega y\right)^{-\frac{\eta v}{\Omega}},
\quad
f_R^{(0)}(y) = 0,
\quad M_0 = 0,
\ee
where $N_{L,0}$ is a normalization constant. 
The number of excited bound KK states corresponds to $n = \lfloor \frac{\eta v}{\Omega} \rfloor$
($\lfloor\,\rfloor$ is the floor function).
For example, the first excited bound state exists when $\frac{\eta v}{\Omega} \ge 1$ and 
its wave function and mass are given by
\be
f^{(1)}_L = N_{L,1} \sinh\Omega y \left(\cosh\Omega y\right)^{- \frac{\eta v}{\Omega}},\quad 
f^{(1)}_R = N_{R,1} Qf^{(1)}_L,\quad 
M_1^2 = \left(2\frac{\eta v}{\Omega} - 1\right)\Omega^2.
\label{eq:F_KK}
\ee
The mass gap between the zero mode and the KK modes is again 
of order $\Omega$ for the parameter region given in Eq.~(\ref{eq:yukawa}).
The analysis for $\tilde \Psi$ can be done similarly by replacing $\eta$ with $\tilde \eta$ and by exchanging
$L$ and $R$.

The interaction between the lightest massive gauge boson $A_\mu^{(0)}$ 
and the fermionic zero mode $\psi_L^{(0)}$ is obtained as 
\be
\int^\infty_{-\infty} dy\, i\bar \Psi \Gamma^\mu {\cal D}_\mu \Psi 
= i \bar \psi_L^{(0)} \gamma^\mu\left(\p_\mu + i q_f 
\frac{\sqrt{2}\mu}{v_h}A_\mu^{(0)}\right)\psi_L^{(0)} + \cdots,
\ee
where the ellipses stand for the massive modes.
Notice the gauge coupling is the same as in Eq.~(\ref{eq:e}). 
We have to emphasize that the effective gauge coupling $e$ is 
the same for any localized fields. The universality is 
ensured by the fact that the wave function of the lightest mode 
of ${\cal A}_\mu$  is always constant.

We can also easily derive an effective Yukawa coupling as follows,
\be
\int^\infty_{-\infty}dy\, \chi {\cal H} \bar\Psi \tilde \Psi \supset
\chi \bar v \tau(b,\tilde b)\bar \psi^{(0)}_L 
\tilde \psi^{(0)}_R,\quad b \equiv \frac{\eta v}{\Omega},\ \tilde b \equiv \frac{\tilde\eta v}{\Omega},
\ee
with a dimensionless constant
$
\tau(b,\tilde b) = \frac{\Gamma(\frac{1+b+\tilde b}{2})}{\Gamma(\frac{2+b+\tilde b}{2})}
\sqrt{
\frac{\Gamma(b+\frac{1}{2})\Gamma(\tilde b + \frac{1}{2})}{\Gamma(b)\Gamma(\tilde b)}
},
$
where $\Gamma(x)$ is the gamma function. Thus the Yukawa coupling in the four dimensions reads
\be
\chi_4 = 
\frac{\tau(b,\tilde b)\chi \bar v}{v_h} \simeq \chi \sqrt\Omega,
\label{eq:yukawa_4}
\ee
where we assume that $\tau(b,\tilde b)$ is of order one because of $b\simeq \tilde b \simeq 1$.

Before closing this section, let us comment on the Higgs field. 
The Higgs condensation occurs at the five-dimensional 
level leading to the localization of the massless/massive gauge 
bosons in our model. 
A new feature of our Higgs mechanism is that the order parameter 
${\cal H}$ induced by domain wall is position-dependent. 
As a consequence, effective Higgs field is localized and only 
the massive physical Higgs boson $h$ remains in the low-energy physics. 
In contrast, if one uses other neutral scalar fields $\phi_i$ 
to localize the gauge fields 
\cite{Arai:2012cx,Arai:2013mwa,Arai:2014hda,Arai:2016jij,
Arai:2017lfv,Arai:2017ntb,Arai:2018rwf}, 
one has to prepare another trick to localize the Higgs fields 
too. For example, in recent papers \cite{Okada:2017omx,Okada:2018von}, 
the kinetic term of the Higgs field is not minimal but 
multiplied by a function $\beta^2(\phi)$. 
In such models, the Higgs field (massive Higgs boson and massless 
NG boson) is localized on the domain wall and the Higgs 
condensation occurs in the low-energy effective theory. 
Namely, the Higgs field plays no active roles at the 
five-dimensional level.

\section{Phenomenological implications}
\label{sec:3}
\subsection{Mass scales}

In order to have a phenomenologically viable model, we 
need to explain observed mass $m_A$ of a gauge boson, vacuum 
expectation value $v_h$ of four-dimensional Higgs field, and 
mass $m_h$ of physical Higgs boson. These observables are 
necessary and sufficient to fix the parameters of gauge-Higgs 
sector of the SM (gauge coupling $e$, and quadratic and quartic 
couplings of Higgs scalars). 
We can regard all these masses to be of order $10^2$ GeV, 
taking the four-dimensional gauge coupling\footnote{
Here we have just one gauge coupling, 
because of our simplification of $U(1)$ instead of $SU(2)\times U(1)$ 
gauge group. } 
 $e$ and Higgs quartic coupling $\lambda_4$ to 
be roughly of order unity\footnote{Actually they are somewhat 
less than unity experimentally, in conformity with the perturbativity 
of SM.}. 
On the other hand, we have four parameters, $\Omega, v, \lambda, \mu$, 
in the bosonic part of the five-dimensional Lagrangian (\ref{eq:Lag}). 
It is convenient to take $\Omega$ as the fundamental mass scale 
of the high energy microscopic theory. 
Three other parameters can be put into two mass scales 
$\mu, \lambda \bar v$, and one dimensionless combination 
$\lambda_4^2 = 2\lambda^2\Omega/3$ in Eq.~(\ref{eq:constant_higgs}),
where $\bar v=\sqrt{v^2-(\Omega/\lambda)^2}$. 
From Eqs.~(\ref{eq:vev}), (\ref{eq:mh}), and (\ref{eq:m_A}), 
masses of the low-energy effective theory are given in terms of 
parameters of the five-dimensional theory as 
\begin{equation}
m_A=\sqrt{2}q_H \mu, 
\quad 
v_h=\frac{2}{\sqrt{\Omega}}\bar v, 
\quad 
m_h=\sqrt{\frac{8}{3}}\lambda \bar v. 
\label{eq:low_energy_masses}
\end{equation}
Fitting these masses to experimentally observed values, we 
still have one mass scale $\Omega$ completely free. 
Therefore we can choose the energy scale $\Omega$ of the 
five-dimensional theory as large as we wish, leaving 
phenomenologically viable model at low-energies.

For instance, if we choose the ratio of the high energy scale 
and SM scale to be parametrized as 
\be
\epsilon^2 = \frac{\lambda^2 \bar v^2}{\Omega^2} \sim 10^{-2a} \ll 1, 
\label{eq:ep2}
\ee
we find the scale of parameters in the model as 
\begin{equation}
\lambda\bar v\sim 10^2\,{\rm Gev} \ll 
\lambda v \sim\Omega\sim 10^{2+a}\,{\rm GeV}, 
\label{eq:fine-tuning}
\end{equation}
implying $\lambda \sim 10^{-1-{a}/{2}}$ GeV${}^{-1/2}$,
$v\sim 10^{3+3a/2}$ GeV${}^{3/2}$, and 
$\bar v \sim 10^{3+a/2}$ GeV${}^{3/2}$. 
This large mass gap allows us to use the low-energy effective 
field theory retaining only light fields with the mass of order 
$\lambda \bar v$ or less. 
In order to achieve this hierarchy, we need a 
fine-tuning of parameters $\lambda \bar v \ll \Omega$, as in 
Eq.(\ref{eq:fine-tuning}).

For the fermionic sector, we require Eqs.~(\ref{eq:yukawa}) and (\ref{eq:yukawa_4}).
Therefore, we have $\eta \sim \tilde \eta \sim 10^{-1-{a}/{2}}$ GeV${}^{-1/2}$.
In order to obtain appropriate values of the four-dimensional Yukawa couplings,
for example, for the top Yukawa coupling to be of order one, we need the five-dimensional Yukawa coupling as
\be
\chi \simeq \frac{\chi_{4,{\rm top}}}{\sqrt\Omega} \sim 10^{-1 -\frac{a}{2}}\, {\rm GeV}^{-\frac{1}{2}}.
\ee
Thus, the five-dimensional Yukawa couplings $\eta$, $\tilde \eta$ and $\chi$ are naturally set to be the same order. Note also that
this justifies Eq.~(\ref{eq:yukawa}). 
To understand the hierarchy of lighter fermion masses, we can 
use the usual mechanism of splitting of position of localized 
fermions as explained briefly in Sec.~\ref{sec:4}.

In summary, for having the SM at the low-energy, 
all the dimension full parameters in the five-dimensional Lagrangian are set to be
of the same order as
\be
\Omega \sim \lambda^{-2} \sim v^{\frac{2}{3}} \sim \eta^{-2} \sim \tilde \eta^{-2} \sim \chi^{-2} \sim 10^{2+a}\, \text{GeV}.
\ee
We need a fine-tuning for two small parameters of mass dimension: 
$\lambda \bar v, \mu \sim 10^2$ GeV. 
Estimate of the lower bound for the parameter $\Omega \sim 10^{2+a}$ GeV 
will be discussed in Sec.~\ref{sec:4} using constraints from the 
LHC data.

\subsection{Translational zero mode}
\label{eq:translational_zero_mode}

Here we study interactions of the translational Nambu-Goldstone 
(NB) mode, and their impact on low-energy phenomenology. 
Symmetry principle gives low-energy theorems, dictating that 
the NG bosons interact with corresponding symmetry currents 
as derivative interactions (no interaction at the vanishing 
momentum of NG bosons). 
Hence their interactions are generally suppressed by powers of 
large mass scale. 
In order to understand the interactions of the NG bosons, it is 
most convenient to consider the moduli approximation\cite{Manton:1981mp} 
where the moduli are promoted to fields in the low-energy effective 
Lagrangian. 
Let us consider a general theory with a number of fields\footnote{
In our concrete model, we have fields such as ${\cal A}_M, 
{\cal T}, {\cal H}$, $\Psi$ and $\tilde \Psi$.} 
$\phi^i(x,y)$ admitting a solution (soliton) 
of field equation, which we take as a background. 
When the theory is translationally invariant, the position $Y$ 
of the soliton is a moduli. It is contained in the solution as 
$\phi^i(x, y-Y)$. 
In the moduli approximation, we promote the moduli parameter $Y$ 
to a field $Y(x)$ slowly varying in the world volume 
of the soliton. We call this moduli field $Y(x)$ as NG field\footnote{
This definition is, in general, a nonlinear field redefinition of 
the effective field that arises in the mode analysis of 
fluctuation fields, such as in Appendix. }. 
By introducing the NG boson decay constant $f_{Y}$ to adjust the 
mass dimension of the NG field to the canonical value $[Y(x)]=1$, 
we obtain 
\begin{equation}
{\cal L}_{\rm NG}=\int dy\, {\cal L}\left(\phi\left(x, 
y-\frac{1}{f_{Y}}Y(x)\right)\right).
\label{eq:NG_action}
\end{equation}
The precise value of the decay constant $f_{Y}$ is determined 
by requiring the kinetic term of NG boson to be canonical as 
illustrated in the subsequent explicit calculation. 
By integrating over $y$, we can obtain the effective interaction 
of the NG field. 
One should note that the constant part $Y$ of NG field $Y(x)$ 
is nothing but the position of the wall, which can be absorbed 
into the integration variable $y$ by a shift $y\to y-Y$ because 
of the translational invariance. Hence the constant
$Y$ disappears from the 
effective action after $y$-integration is done. 
This fact guarantees that $Y(x)$ must appear in the low-energy 
effective theory always with derivatives, i.e. $\partial_\mu Y(x)$. 
Let us examine how this fact fixes the interactions of NG particle 
in the effective Lagrangian to produce the low-energy theorem. 
Derivative $\partial_\mu$ can only come from the derivative 
term in the original action ${\cal L}$, giving terms linear 
in the NG particle $Y(x)$ as 
\begin{equation}
{\cal L}_{\rm NG}
=-\int dy\, \frac{\partial {\cal L}}{\partial \partial_\mu \phi^i}
\frac{\partial \phi^i}{\partial y}\frac{\partial_\mu Y}{f_{Y}}+\cdots
=-\frac{1}{f_{Y}}\partial_\mu Y(x) \left[\int dy\, T^{\mu y}\right] +\cdots ,
\label{eq:NG_int_linear}
\end{equation}
where the energy-momentum tensor $T^{MN}$ of matter in five 
dimensions is given by  
\begin{equation}
T^{MN}=\frac{\partial {\cal L}}{\partial \partial_M \phi^i}
\partial^N \phi^i-\eta^{MN}{\cal L} .
\label{eq:EMtensor}
\end{equation}
This is the low-energy theorem of the NG particle for spontaneously 
broken translation. 
Thus we find that there are no nonderivative interactions that 
remain at the vanishing momentum of NG bosons, including KK particles. 
For instance, the possible decay amplitude of a KK fermion into 
an ordinary fermion and a NG boson should vanish at zero momentum 
of NG boson and will be suppressed by inverse powers of large 
mass scale such as $\Omega$. 
In this way, we can compute the effective action of NG field 
in powers of derivative $\partial_\mu$. 
Usually we retain up to second order in derivatives, but 
higher derivative corrections can be obtained systematically 
with some labor~\cite{Eto:2012qda}.

Let us compute the effective Lagrangian of NG field 
$Y(x)$ more explicitly by using the moduli approximation 
in our model as 
\begin{equation}
{\cal T} = v \tanh\left(\Omega y - \frac{1}{f_Y}
Y(x)\right), \hspace{5mm}
{\cal H} = \sqrt{\frac{\Omega}{2}}H(x)\,\sech\left(\Omega y 
- \frac{1}{f_Y}Y(x)\right). 
\end{equation}
The wall position moduli in wave functions of fermions 
must also be promoted to NG field $Y(x)$, i.e. 
\begin{equation}
f_{L,R}^{(n)}(y) \to f_{L,R}^{(n)}\left(y
-\frac{1}{f_Y}Y(x)\right),
\end{equation}
although we only retain the zero mode given in 
Eq.~(\ref{eq:fermzeromode}) in order to obtain low-energy 
effective Lagrangian for light particles. 
Plugging these Ansatz into the four-dimensional kinetic 
terms of ${\cal T}$, ${\cal H}$ and $\Psi$ and integrating 
over $y$, we obtain the effective Lagrangian 
containing the NG field. Requirement of canonical normalization 
of the NG field $Y(x)$ fixes the decay constant $f_Y$ as 
\begin{equation}
f_Y=\frac{2\sqrt{2}\, v}{\sqrt{3\Omega}}. 
\label{eq:decay_const}
\end{equation}
We finally obtain the effective Lagrangian for low-energy particles as: 
\begin{equation}
{\cal L}_{\rm NG} = \frac{1}{2}\partial_\mu Y \partial^\mu Y 
\left(1+\frac{\Omega}{2v^2}|H|^2\right).
\label{eq:NG_action}
\end{equation}
A few features can be noted. 
First of all, the NG bosons have only derivative interactions, 
as required by the above general consideration. 
Secondly, the derivative interaction produces higher-dimensional 
operators coupled to NG bosons. 
The required mass parameter in the coefficient of the interaction 
term is given by the high energy scale as 
${\Omega}/({2v^2})\sim 1/\Omega^2$.
Therefore the interaction is suppressed by a factor of 
(momentum)$/\Omega$. 
Thirdly, the interaction linear in the NG particle in 
Eq.(\ref{eq:NG_int_linear}) happens to be absent in this model. 
This is a result of a selection rule in our model.\footnote{
Note that 
a non-derivative coupling $Y\bar\psi_L^{(0)} \psi_R^{(1)}$
from ${\cal T}\bar\Psi\Psi$  was recently studied in 
Ref.~\cite{Okada:2018von}. 
However, the symmetry principle of NG boson for translation does 
not allow coupling without the derivative $\p_\mu$.} 
The Lagrangian (\ref{eq:Lag}) and the background solution 
(\ref{eq:T1}) allows us to assign generalized parity 
under the reflection symmetries $y \to -y$, as a conserved 
quantum number to all modes including KK modes. 
Since NG boson has odd parity, whereas all other low-energy 
particles including fermion have even parity, 
we end up in the quadratic interaction for the NG boson $Y(x)$, 
as given in Eq.~(\ref{eq:NG_action}). 
The parity quantum number under $y\to -y$ may not be  
conserved in more general models, and can have nonvanishing 
interaction linear in $\partial_\mu Y(x)$ given in 
Eq.~(\ref{eq:NG_int_linear}).

Only when we take into account the heavy KK modes \cite{Okada:2018von}, 
we have interactions linear in $\p_\mu Y$.
For example, including the lightest KK fermion given in Eq.~(\ref{eq:F_KK}) 
($b = \frac{\eta v}{\Omega}>1$ in order to have a discrete state) 
we obtain a vertex 
\be
\int^\infty_{-\infty}dy\, i \bar \Psi \Gamma_M {\cal D}^M \Psi \supset 
i \alpha \frac{\sqrt\Omega}{v} \p_\mu Y\left(
\bar\psi^{(1)}_L \gamma^\mu\psi^{(0)}_L - 
\bar\psi^{(0)}_L \gamma^\mu\psi^{(1)}_L\right), 
\ee
where $\alpha$ is a dimensionless constant of order one defined by 
$
\alpha \equiv  \frac{\sqrt3}{4}\frac{b}{\sqrt{b-1}}\frac{B(b
+\frac{1}{2},b-\frac{1}{2})}{B(b+1,b-1)}\,,
$ where $B(x,y)$ is the beta function.\footnote{Note that $\alpha \to 0$ as $b\to1$.}
The above interaction gives the decay process 
$\psi^{(1)}_L \to Y\psi^{(0)}_L$.
$\tilde \Psi$ yields similar interactions between $Y$ and $\psi_R^{(n)}$.
Although the NG boson amplitudes are generally suppressed 
by the ratio  $p_\mu/\Omega$ with the large mass scale $\Omega$, 
it can give a significant decay rate in the case of two-body 
decay like here. 
Moreover, this type of vertex provides a new diagram for the
production of KK quarks $\psi_{L,R}^{(1)}$ out of quarks $\psi_{L,R}^{(0)}$ 
in the colliding nucleons via the NG boson exchange 
\begin{eqnarray}
\psi_i^{(0)}+\psi_j^{(0)}\to \psi_i^{(1)}+\psi_j^{(1)}, \quad (i,j=L,R) \label{eq:NGexchange}
\end{eqnarray}
in the LHC experiment. 
This should be the dominant production mechanism 
because of large momentum fraction of quarks as given by their 
distribution function inside nucleons. 
The production process (\ref{eq:NGexchange}) tells us the lower 
bound of the KK quark masses. We will estimate it in Sec.~\ref{sec:4} where
the Standard Model is embedded in our framework.

\subsection{$h\to \gamma\gamma$}

As explained above, our model provides a domain wall inside 
which all the SM particles are localized.
All the KK modes are separated by the mass gap $\Omega \sim 10^{2+a}$ GeV. 
Furthermore, for the minimal $\beta^2$ as given in 
Eq.~(\ref{eq:beta_simple}), there are no additional localized KK modes of 
the gauge fields \cite{Arai:2018rwf}.
At first sight, one might wonder if the low-energy theory would 
be distinguishable from the conventional SM if $a$ is 
sufficiently large. 
However, a significant difference between these two theories is 
an additional interaction between the Higgs boson and the gauge 
bosons due to the field dependent gauge kinetic term.
For illustration, suppose that ${\cal A}_M$ is the electromagnetic 
gauge field and the Higgs boson is neutral with $q_H=0$.
Nevertheless, the field-dependent gauge kinetic term yields 
an interaction between the photon and the neutral Higgs boson. 
This mechanism is valid also for the physical Higgs boson in 
our model. To see this, let us consider the fluctuation of 
physical Higgs boson $h(x)$ by perturbing $H$ in Eq.~(\ref{eq:H}) 
about $H= v_h$ 
\be
{\cal H} = \bar v \left(1 +  \frac{\sqrt{2}h(x)}{v_h}
\right)\, \sech\, \Omega y.
\label{eq:H_fluc}
\ee
Then the first term of Eq.~(\ref{eq:Lag}) yields
\be
- \int_{-\infty}^\infty dy\, |\beta|^2 ({\cal F}_{MN})^2  
= - \frac{1}{4} \left(1 + 2\frac{\sqrt{2}h}{v_h} 
+ \frac{2h^2}{v_h^2}\right) (F_{\mu\nu}^{(0)})^2.
\label{eq:hgg}
\ee
Thus, there is a new tree-level amplitude for $h \to \gamma\gamma$. 
In the SM, the Higgs boson decays into two photons mediated 
by top or $W$ bosons at one-loop level. 
The operator of interest is $c\frac{h}{v_h} (F_{\mu\nu}^{(0)})^2$, 
whose coefficient is bounded by the LHC measurement as 
$c \sim 10^{-3}$ \cite{Ellis:2014dva,Ellis:2014jta}. 
However, our simplest model has $c = \frac{1}{2}$, 
so is strongly excluded experimentally.

\subsection{Generalized models}
\label{sec:genmod}

To have a phenomenologically acceptable $h\to\gamma\gamma$ decay 
amplitude, we can modify the field dependent gauge kinetic term, 
for example,  as 
\be
\beta^2({\cal H}) =  \frac{1}{2\mu^2}\left(|{\cal H}|^2 
- \frac{3}{4}\frac{|{\cal H}|^4}{\bar v^2}\right).
\label{eq:beta_mod}
\ee
The background configuration of the Higgs field 
${\cal H} = {\cal H}_0(y)$ remains the same as in Eq.~(\ref{eq:T1})
since the $\beta^2 {\cal F}_{MN}^2$ term does not contribute 
to the background solution. 
The reason for selecting this specific modification will be 
explained below soon. Before that, however, let us mention 
that the modification comes with a price. 
The linearized equation of motion in the generalized $R_\xi$ gauge 
for the gauge field with a generic $\beta$ reads 
\cite{Arai:2018rwf,Arai:2018xyz}
\be
\left\{\eta^{\mu\nu}\square - \left(1-\frac{1}{\xi}\right)\p^\mu\p^\nu 
+ \eta^{\mu\nu}\left(-\p_y^2 + \frac{(\p_y^2\beta)}{\beta} 
+ q_H^2\frac{{\cal H}_0^2}{2\beta^2}\right)\right\}A_\nu = 0.
\ee
Then, determining the physical spectrum corresponds to solving 
the eigenvalue problem 
\be
\left(-\p_y^2 + \frac{(\p_y^2\beta)}{\beta} 
+ q_H^2\frac{{\cal H}_0^2}{2\beta^2}\right) \phi_n = m_n^2 \phi_n.
\ee
If $\beta^2$ is quadratic in ${\cal H}$ as was the case in 
Eq.~(\ref{eq:beta_simple}), the third term on the left-hand side 
is constant. 
Therefore, the problem is of the same complexity as if $q_H=0$. 
On the other hand, when $\beta^2$ is not purely quadratic, the 
eigenvalue problem is essentially different from that of 
$-\p_y^2 + \frac{(\p_y^2\beta)}{\beta}$.
Fig.~\ref{fig:fig} (b) shows the corresponding Schr\"odinger potential.
In case of (\ref{eq:beta_mod}) the Schr\"odinger equation in 
terms of the dimensionless coordinate $z = \Omega y$ is given by
\be
\left(- \p_z^2  + \frac{\p_z^2 \beta_0}{\beta_0} + q_H^2 
\frac{\mu^2}{\Omega^2} \frac{2}{1 
- \frac{3}{4}\frac{{\cal H}_0^2}{\bar v^2} }\right)
\phi_n = \frac{m_n^2}{\Omega^2} \phi_n.\quad \beta_0 = \beta({\cal H}_0).
\ee
Note that this is independent of $\bar v$ because of 
${\cal H}_0 = \bar v\, \sech\, z$. 
Although we cannot solve this exactly, we can still solve this 
problem perturbatively for $\Omega \gg \mu$ by treating the 
third term on the left-hand side as a small correction. 
The lowest eigenfunction and eigenvalue are approximately given by
\be
\phi_0 = \frac{\mu \sqrt{2\Omega}}{\bar v} \beta_0,\quad 
m_0^2 \simeq 2 q_H^2 \mu^2 
\int dy\, \phi_0^2\left(1-\frac{3}{4}\, \sech^2\Omega y\right)^{-1} 
= 2q_H^2 \mu^2.
\label{eq:phi0_approx}
\ee
This is just the same as Eq.~(\ref{eq:phi_0}), and, therefore, 
the mass of the lightest massive gauge boson is of order $\mu$, which
justifies our assumption $\Omega \gg \mu$.
Since the situation is almost the same as in the simplest model, 
we have $v_h^2/2 = \int dy\, {\cal H}_0^2 = 2\bar v^2/\Omega$, 
and the effective gauge coupling is $e \sim \mu/v_h \sim 1$.
Thus the modified model defined by Eq.~(\ref{eq:beta_mod}) 
provides the SM at low energies in the same manner 
as the simplest model does.

Now, let us turn to the problem of $h \to \gamma \gamma$. 
So we set $q_H=0$ and Eq.~(\ref{eq:phi0_approx}) becomes exact 
wave function of the massless photon.
As before, we put ${\cal H}$ given in Eq.~(\ref{eq:H_fluc}) into 
the gauge kinetic term $-\beta^2 {\cal F}_{MN}^2$ with $\beta^2$ 
given in Eq.~(\ref{eq:beta_mod}). Then, we find 
\be
- \int^\infty_{-\infty}dy\, \beta^2 ({\cal F}_{\mu\nu})^2 
=  \left[- \frac{1}{4} + \frac{2h^2}{v_h^2}  
+ {\cal O}\left(\frac{h^3}{v_h^3}\right)\right] (F_{\mu\nu}^{(0)})^2 .
\label{eq:higgs_gauge_int}
\ee
As we see, the term $h (F_{\mu\nu}^{(0)})^2$ does not exist. 
Therefore, the modified model is compatible with the bound given 
by the current experimental measurement of $h \to \gamma\gamma$.

If the factor in front of the quartic term of Eq.~(\ref{eq:beta_mod}) 
deviates slightly from $\frac{3}{4}$, the term 
$h (F_{\mu\nu}^{(0)})^2$ comes back with a tiny factor. 
We can compare the contribution of this tree-level term to 
$h\to \gamma\gamma$ with those mediated by top/$W$-boson loop 
in the SM. 
If a sizable discrepancy is found in the future experiments 
in $h\to\gamma\gamma$ channel compared with the SM prediction, 
it can be a signature of our model. 

Of course, the modification in Eq.~(\ref{eq:beta_mod}) is just 
an example. There are other modifications which 
forbid $h\to\gamma\gamma$ process at the tree-level.
For instance, in addition to $h\gamma\gamma$, one can eliminate 
other higher-dimensional interactions such as $hh\gamma\gamma$ 
vertex by appropriately choosing $\beta^2$.

The above consideration holds for another similar process 
of $h\to gg$ (two gluons). An experimental signature should be the decay 
of physical Higgs particle to hadronic jets. 
Moreover, it will affect the production rate of physical 
Higgs particles from hadron collisions. 

Recently, it was proposed that another interesting signature 
from the localized heavy KK 
modes of  gauge bosons and fermions \cite{Okada:2017omx,Okada:2018von}, 
although the presence and/or the number of localized KK modes 
is more dependent on details of models. 
Our model has the same signatures too but they are subdominant in our model since
they are 1-loop effects of the supermassive KK modes.

\section{The Standard Model}
\label{sec:4}

Let us briefly describe how our mechanism works in the SM. 
The minimal five-dimensional Lagrangian is
\be
{\cal L} &=& - \beta({\cal H})^2 \left[\left({\cal G}_{MN}^a\right)^2 
+ \left({\cal W}_{MN}^i\right)^2 + {\cal B}_{MN}^2\right] 
+ \left|{\cal D}_M{\cal H}\right|^2 + (\p_M {\cal T})^2  - V \nonumber\\
&+& i \bar U\Gamma^M{\cal D}_MU + i \bar Q \Gamma^M{\cal D}_M Q 
+ \eta_R \left({\cal T}-m\right)\bar U U - \eta_L {\cal T} \bar Q Q + \chi  \bar Q {\cal H} U 
+ {\rm h.c.}\,,
\label{eq:lag_SM}\\
V &=& \Omega^2 |{\cal H}|^2 + \lambda^2 \left({\cal T}^2 
+ |{\cal H}|^2 - v^2\right)^2,
\ee
where ${\cal G}_{MN}$, ${\cal W}_{MN}$, and ${\cal B}_{MN}$ are 
the field strengths of $SU(3)_C$, $SU(2)_W$ and $U(1)_Y$ 
gauge fields, respectively. More explicitly, they are given by
${\cal W}_{MN} = \p_M {\cal W}_N - \p_N {\cal W}_M + i q \left[{\cal W}_M,{\cal W}_N\right]$,
and so on.
The Higgs field ${\cal H}$ is an $SU(2)_W$ doublet with the 
covariant derivative 
${\cal D}_M {\cal H} = \left(\p_M + \frac{i}{2}q {\cal W}_M 
+ \frac{i}{2}q'{\cal B}_M\right) {\cal H}$,
with $q$ and $q'$ being five-dimensional  gauge couplings 
for $SU(2)_{W}$ and $U(1)_Y$ relative to that of $SU(3)_C$.  
We will assume $\tan\theta_{\rm w} = q'/q$
to reproduce the SM in the low-energy.
The fermions $Q$ and $U$ are doublet and singlet of $SU(2)_W$, 
respectively. Flavor indices for $U$, $Q$ and the couplings are implicit.

As before, there are two discrete vacua ${\cal T}= \pm v$ and 
${\cal H} = 0$. The background domain wall solution in the 
parameter region $\lambda v > \Omega$ is given by 
\be
{\cal T}_0 = v \tanh \Omega y,\quad
{\cal H}_0 = \left(
\begin{array}{c}
0 \\
\bar v \, \sech\,\Omega y
\end{array}
\right).
\ee
The Higgs doublet $H(x)$ in the four-dimensional effective 
theory is found in ${\cal H}$ as is done in Eq.~(\ref{eq:H}).  
The Higgs potential is identical to that in Eq.~(\ref{eq:L_higgs}).
One can show that the upper component and the imaginary part 
of the lower component are localized NG bosons and are absorbed 
by the $W$ and $Z$ bosons. Indeed, the spectrum of 
$W_\mu^{\pm} = 2\beta {\cal W}_\mu$, and 
$Z_\mu = 2 \beta {\cal Z}_\mu$ are determined by the 1D 
Schr\"odinger problems 
\be
-\p_y^2 + \frac{(\p_y^2\beta)}{\beta} + \frac{q^2}{4} 
\frac{{\cal H}_0^2}{2\beta^2},\quad 
-\p_y^2 + \frac{(\p_y^2\beta)}{\beta} + \frac{q^2}{4\cos^2\theta_{\rm w}} 
\frac{{\cal H}_0^2}{2\beta^2}.
\label{eq:WZ}
\ee
The details of the derivation 
will be given elsewhere \cite{Arai:2018xyz}.
On the other hand, the photon $A_\mu = 2\beta {\cal A}_\mu$ and 
gluon $G_\mu = 2\beta {\cal G}_\mu$ are determined by 
$-\p_y^2 + \frac{(\p_y^2\beta)}{\beta}$.
Therefore, the lightest modes $\phi_0 \propto \beta$ of photon 
and gluon are exactly massless. 
The results so far are independent of $\beta^2$. 
To be concrete, let us choose the simplest function 
$\beta^2 = |{\cal H}|^2/4\mu^2$. 
Then the effective $SU(2)_W$ gauge couplings and the electric 
charge are given by 
\be
g = \frac{\sqrt{2}q\mu}{v_h},\quad g' = \frac{\sqrt{2}q'\mu}{v_h},\quad
e = \frac{qq'}{\sqrt{q^2 + q'{}^2}}\frac{\sqrt{2}\mu}{v_h} 
= \frac{gg'}{\sqrt{g^2+g'{}^2}}\,,
\ee
where $v_h$ is given in Eq.~(\ref{eq:vev}). 
Masses of $W$ and $Z$ are easily read from Eq.~(\ref{eq:WZ}) as 
\be
m_W^2 = \frac{q^2\mu^2}{2} = \frac{g^2 v_h^2}{4},\quad
m_Z^2 = \frac{q^2\mu^2}{2\cos^2\theta_{\rm w}} = 
\frac{g^2 v_h^2}{4\cos^2\theta_{\rm w}}.
\ee

For the fermions, we assume $\eta_L > 0$ and $\eta_R > 0$. 
Then the left-handed fermion from $Q$ is localized at the zero of ${\cal T}$, while 
the right-handed 
fermion from $U$ is localized at the zero of ${\cal T} - m$. 
The Yukawa term $\chi\bar Q{\cal H} U$ is responsible 
for giving non-zero masses to the localized chiral fermions, 
which is necessarily exponentially small for $m\neq0$ since the left- and 
right-handed fermions are split in space. 
By distinguishing parameters such as $m$ 
for different generations as was done in many models with extra 
dimensions~\cite{ArkaniHamed:1999dc,Dvali:2000ha}, the hierarchical 
Yukawa coupling can be naturally explained in our model.

This way, the SM particles are correctly localized on the domain 
wall in our framework.

\begin{figure}[b]
\begin{center}
\includegraphics[width=10cm]{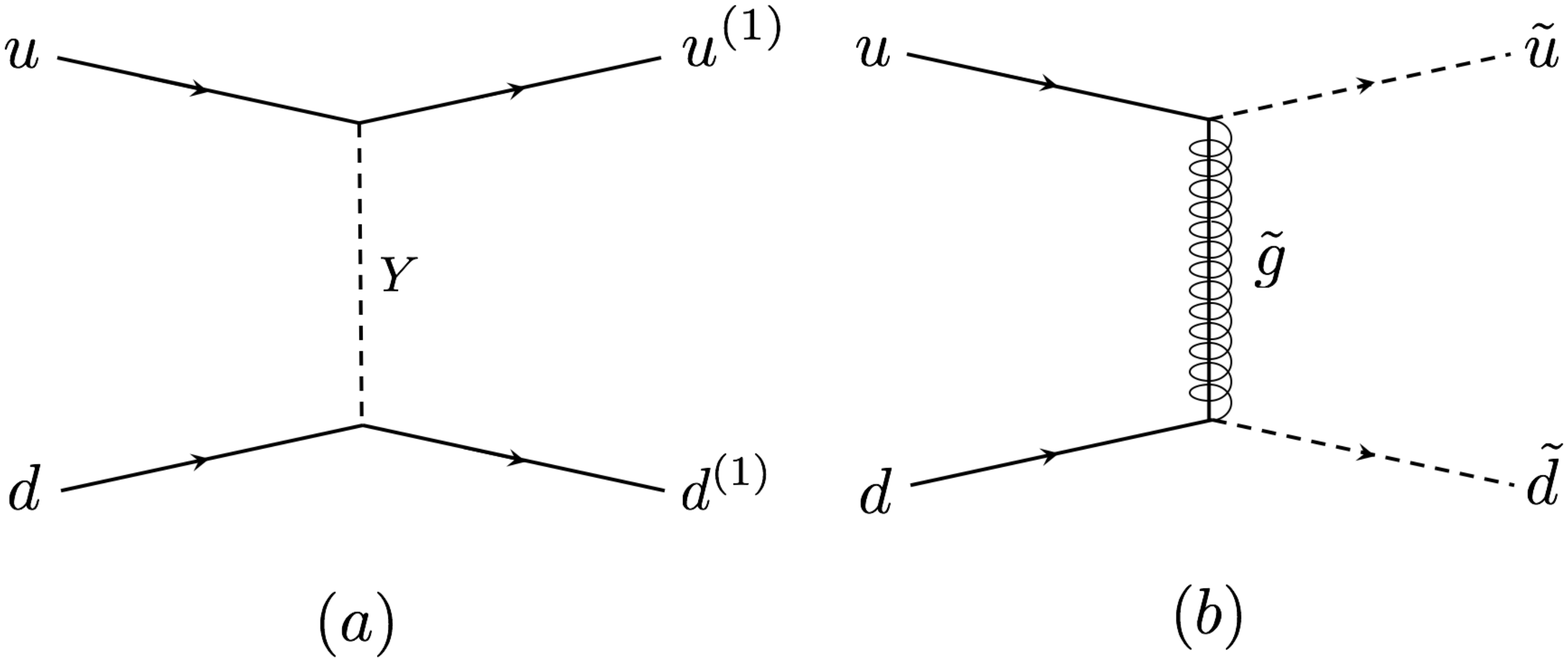}
\caption{Feynman diagrams for the processes $(a)~ud \rightarrow u^{(1)}d^{(1)}$ and $(b)~ud \rightarrow \tilde{u}\tilde{d}$.}
\label{fig:diagram}
\end{center}
\end{figure}
Before closing, we evaluate the lower bound of KK quark mass by using 
the KK quark production process in Eq.~(\ref{eq:NGexchange}) 
via Nambu-Goldstone boson exchange.
If we take the initial quarks of different flavor for simplicity, 
we have only single Feynman diagram depicted in Fig.~\ref{fig:diagram}(a). 
In the process (\ref{eq:NGexchange})  
followed by $\psi_{L,R}^{(1)} \to Y\psi_{L,R}^{(0)}$,
the final state contains two SM fermion jets 
and a missing energy of the NG boson $Y$, whose signature is 
similar to squark pair production, where a squark decays into the 
partner SM quark and a gluino or neutralino in the simplified 
supersymmetric models \cite{Alwall:2008va,Alwall:2008ag,Alves:2011wf}. 
In most of kinematical regions, a dominant processes for squark 
pair production is given by Feynman diagram depicted in 
Fig.~\ref{fig:diagram}(b). 
Since both processes involve the same valence quark distribution 
functions, we can compare these cross-sections directly to obtain 
an order of magnitude estimate of the lower bound for KK quark mass 
using the analysis for squark mass bound. 
As shown in Appendix \ref{sc:cross}, the differential cross 
section ${d\sigma \over dt}$ of (\ref{eq:NGexchange}) producing 
a pair of the first KK fermion with mass $M_1$ is given by 
summing contributons from initial state of different chiralities 
($LL, RR, LR, RL$) as 
\begin{eqnarray}
{d\sigma \over dt}(u d\rightarrow u^{(1)}d^{(1)})
= {\alpha^4 \over 576
 \pi^2 
}{\Omega^2 \over v^4}
{
(1-\beta_{M_1}\cos\theta)^2(1-\beta_{M_1}^2) \over 
(\beta_{M_1}^2+1-2\beta_{M_1}\cos\theta)^2},
\label{eq:cross1}
\end{eqnarray}
where $\beta_{M_1}=\sqrt{1-{M_1^2 \over E^2}}$, 
$E$ is the center of mass energy of incoming particles, and 
$\theta$ is the scattering angle. We ignore masses of the SM quarks and all the parameters
are taken to be common for the different quarks just for simplicity.
We can assume $v \approx \Omega^{3/2}$ and $M_1\approx \Omega$ 
for simplicity.

The squarks production $ud\rightarrow \tilde{u}\tilde{d}$ cross 
section \cite{Harrison:1982yi, Dawson:1983fw, Reya:1984yz} is 
\begin{eqnarray}
 {d\sigma \over dt}(ud\rightarrow \tilde{u}\tilde{d})={g_s^4 \over 288
\pi 
}
 {1+\beta_{m_{\tilde{q}}}^2\cos^2\theta
+(m_{\tilde{g}^2}-m_{\tilde{q}}^2)/E^2 \over 
(2E^2(1-\beta_{m_{\tilde{q}}}\cos\theta)+m_{\tilde{g}}^2-m_{\tilde{q}}^2)^2}, 
\label{eq:cross2}
\end{eqnarray}
with $\beta_{m_{\tilde{q}}}=\sqrt{1-{m_{\tilde{q}}^2 \over E^2}}$. 
The $SU(3)_C$ gauge coupling and gluino mass are denoted as $g_s$ 
and $m_{\tilde{g}}$, and a common mass $m_{\tilde{q}}$ is assumed 
for squarks of different flavors and chiralities. 

To obtain the bound for the production of heavy particles, 
we can expect that the cross-section near threshold ($\beta=0$) is a good 
guide for the order of magnitude estimate. 
Both differential cross-sections become constants without 
angular dependence at the threshold, and their ratio is given as 
\begin{equation}
\frac{{d\sigma \over dt}(u d\rightarrow u^{(1)}d^{(1)})|_{E=M_1}}
{{d\sigma \over dt}(ud\rightarrow \tilde{u}\tilde{d})|_{E=m_{\tilde{q}}}} 
=\frac{1}{2\pi}\frac{\alpha^4}{g_s^4}\frac{m_{\tilde{g}}^2m_{\tilde{q}}^2}
{\Omega^4}\left(1+\frac{m_{\tilde{q}}^2}{m_{\tilde{g}}^2}\right)^2. 
\end{equation}
The simplified analysis for squark production gives $m_{\tilde{q}}> 1.5~$TeV,
assuming
$m_{\tilde{q}}=m_{\tilde{g}}$ \cite{bound}. 
The identical bound for the KK fermion mass $M_1\sim \Omega > 1.5$~TeV 
is obtained for $2\alpha^4/(\pi g_s^4)\approx 1$. 
Since $\Omega =10^{2+a}$~GeV, we have the lower bound for $a$ as $a\gtrsim 1$.
If the coupling $\alpha$ of KK fermion is larger than $g_s$, we obtain 
larger lower bound for its mass. 
To determine how much larger requires a more detailed analysis of data.

\section{Finite electroweak monopoles}
\label{sec:5}

The SM has a point magnetic monopole which is the so-called 
Cho-Maison (CM) monopole \cite{Cho:1996qd}. It is different 
from either a Dirac monopole or a Nambu electroweak monopole 
\cite{Nambu:1977ag}. 
Unfortunately, its mass diverges due to a singularity at the 
center of the monopole. 
Cho, Kim and Yoon (CKY) \cite{Cho:2013vba} have proposed a 
modification of the SM in four dimensions which includes the 
field dependent gauge kinetic term as 
${\cal L} \in - \frac{\epsilon(|H|/v_h)}{4} (B_{\mu\nu})^2$.
In order to have the conventional SM at the electroweak vacuum 
$|H| = v_h$, the normalization should be fixed as 
$\epsilon(|H| \to v_h) = 1$.
It was found that this modification makes the CM monopole regular 
if $\epsilon \sim |H|^n$ with $n > 4+2\sqrt3 \simeq 7.46$ as $|H| \to 0$.
However, it has recently been pointed out by Ellis, Mavromatos 
and You (EMY) \cite{Ellis:2016glu} that the original CKY model 
is incompatible with LHC measurements of Higgs boson 
$H\to\gamma\gamma$. They have proposed generalizations of the 
CKY model which are compatible with LHC measurements.
Their conclusion is that the monopole mass is $\lesssim 5.5$ TeV
so that it could be pair-produced at the LHC and accessible 
to the MoEDAL experiment \cite{Acharya:2016ukt,Acharya:2017cio}.

Neither CKY nor EMY does discuss the underlying rationale for 
their modifications to the SM. 
In contrast, our five-dimensional model has a clear motivation 
for the field dependent gauge kinetic term, which is the domain 
wall induced Higgs mechanism. 
For example, one of the EMY's proposals is \cite{Ellis:2016glu}
\be
\epsilon_1 = 5 \left(\frac{H}{v_h}\right)^8 
- 4 \left(\frac{H}{v_h}\right)^{10}.
\label{eq:epsilon_EMY}
\ee
This can be derived from our model with 
\be
\beta^2 = \frac{|{\cal H}|^2}{\mu^2}\left(10 
\frac{|{\cal H}|^6}{\bar v^6} -9 \frac{|{\cal H}|^8}{\bar v^8}\right).
\label{eq:beta2_EMY}
\ee
The background solution is still ${\cal H}_0 = \bar v\, \sech\, \Omega y$.
Fig.~\ref{fig:fig} (c) shows the corresponding Schr\"odinger potential.
Then the wave function of the massive $U(1)_Y$ gauge field reads 
$\phi_0 \simeq \sqrt{\frac{35\Omega}{64 \bar v^8}
\left(10 {\cal H}_0^8 - 9 \frac{{\cal H}_0^{10}}{\bar v^2}\right)}$\,. 
As before, we identify the four-dimensional Higgs field $H(x)$ as 
${\cal H} = \bar v \frac{H(x)}{v_h}\, \sech\,\Omega y$
with $v_h = \sqrt{\frac{2}{\Omega}}\, \bar v$.
We find the EMY's model from the five dimensions via the domain 
wall and the Higgs mechanism as 
\be
- \int^\infty_{-\infty}dy\, \beta^2 ({\cal B}_{\mu\nu})^2 = 
- \int^\infty_{-\infty}dy\, \beta^2 \frac{\phi_0^2}{4\beta_0^2}
(B^{(0)}_{\mu\nu})^2 = - \frac{\epsilon_1}{4}
(B^{(0)}_{\mu\nu})^2,
\ee
where we ignored contributions from the massive KK modes.

Note that the $\beta^2$ modifies not only the gauge kinetic 
term of $U(1)_Y$ but also that of the $SU(2)_W$.
An electroweak monopole in such theory also has a finite mass 
\cite{Blaschke:2017pym}. 

CKY have claimed that discovery of an electroweak monopole is 
a real final test for the SM \cite{Cho:2013vba} .
For us, it is not only the topological test of the SM but also 
would give constraints for restricting the $\beta^2$ factor
of the five-dimensional theory.

\section{Conclusions and discussion}
\label{sec:6}

We proposed a minimal model in flat non-compact five dimensions 
which realizes the SM on a domain wall. 
In our approach, the key ingredients for achieving this result 
are the following: (i) the spacetime is five-dimensional,
(ii) there is an extra scalar field ${\cal T}$ which is 
responsible for the domain wall, (iii) there is a field-dependent 
gauge kinetic term as a function of the absolute square of 
the Higgs field.

In our model, all spatial dimensions are treated on the same 
footing at the beginning. The effective compactification of 
the fifth dimension happens as a result of the domain 
wall formation breaking the $Z_2$ symmetry spontaneously. 
The presence of domain wall automatically localizes chiral fermions  
\cite{Jackiw:1975fn,Rubakov:1983bb}. The key feature of our 
model is that the Higgs dependent gauge kinetic term drives 
the localization of SM gauge bosons and the electroweak symmetry 
breakdown \emph{at the same time}. The condensation of the SM 
Higgs field inside the wall for $\Omega<\lambda v$ can be 
understood as follows. 
As we let the parameter $\Omega$ decrease across $\lambda v$, 
we find a massless mode emerges at the critical point 
$\Omega=\lambda v$, which becomes tachyonic below the critical 
point and condenses until a new stable configuration 
is formed. 
It is interesting to observe that this thought-process is 
analogous to a second-order phase transition if we regard 
the parameter $\Omega$ as temperature.

Contrary to the conventional wisdom in domain wall model-building, 
where the formation of the domain wall happens separately from
the Higgs condensation to break 
electroweak symmetry, we succeeded in our model to combine both 
mechanisms 
and keep the Higgs field active even in five dimensions. 
In other words, our model is very economical in terms of field 
content. 
Naively, one may expect that this means that domain-wall mass 
scale must coincide with the SM scale, but surprisingly, that 
does not have to be so. As we have argued in Sec.~\ref{sec:3} 
all light modes are separated from all KK modes by the mass 
scale $\Omega$, which is of order $10^{2+a}$~GeV, where 
$a$ can be large at the cost of only mild fine-tuning.
We found a natural bound $a \gtrsim 1$ in Sec.~\ref{sec:4}. The 
reason why this separation of scales happens naturally is that 
we are near the critical point of the domain-wall induced 
Higgs condensation. 
In short, our model can be viewed as an enrichment of the 
conventional domain wall model-building toolbox by a new 
instrument, which is the domain-wall induced condensation 
where the Higgs field plays a role of a position-dependent 
order parameter.

In addition to the conceptual advantages listed above, we 
investigated a new interaction $h\gamma\gamma$ (and $hgg$) coming from 
Eq.~(\ref{eq:f2}). 
This should be bounded by the LHC measurement 
\cite{Ellis:2014dva,Ellis:2014jta}, therefore 
it gives a constraint to $\beta^2$. 
However, a small deviation from exactly vanishing amplitude 
$h\gamma\gamma$ from tree-level coupling is allowed, which 
can be a testable signature in the future experiment at the 
LHC. 
This possibility of the tree-level coupling of $h\gamma\gamma$ 
is a new signature of our model of domain-wall-induced Higgs 
condensation and gauge field localization. 
This feature is in contrast to similar models of gauge field 
localization without the active participation of Higgs field 
in the localization mechanism \cite{Okada:2017omx,Okada:2018von}. 
For instance, these models generally give only loop-effects of 
KK particles, instead of the tree-level $h\gamma\gamma$ coupling. 
Therefore we can have a testable signature of $h\gamma\gamma$ 
even if there are no low-lying KK particles, unlike these models. 
Furthermore, our five-dimensional model explains higher 
dimensional interaction as Eq.~(\ref{eq:epsilon_EMY}) that 
allows the existence of a finite electroweak monopole, whereas 
previous studies have failed to provide the origin of such 
higher-dimensional operators \cite{Cho:2013vba,Ellis:2016glu}. 
The monopole mass was estimated \cite{Cho:2013vba,Ellis:2016glu} as $\lesssim 5.5$ TeV, so that 
it can be pair-produced at the LHC and accessible 
to the MoEDAL experiment \cite{Acharya:2016ukt,Acharya:2017cio}. 
If an electroweak monopole will be found, it provides 
an indirect evidence for the extra dimensions and the domain wall. 
Our domain wall model can account for the hierarchical Yukawa coupling in the
SM from position difference of localized wave functions of matters 
as was done in many models with extra dimensions~\cite{ArkaniHamed:1999dc,Dvali:2000ha}.

If we introduce the other scalar fields $\phi_i$ to localize the 
gauge field and the Higgs field via $\beta(\phi_i)$  as in 
Eq.~(\ref{eq:f2}), they would give an impact on the low-energy 
physics like $\phi_i \to hh$, 
$\phi_i \to \gamma\gamma$, and $\phi_i \to gg$. Therefore, we 
have to be very cautious for including the extra scalar fields $\phi_i$.
Our model is free from this kind of concern, which is 
one of the important progress achieved in this work.

Although we did not explain it in detail, 
the absence of additional light scalar boson from $A_y$ is one 
of the important properties of our model \cite{Okada:2017omx,
Arai:2018rwf,Okada:2018von}.
Moreover, the fact that the localization of gauge fields via 
Eq.~(\ref{eq:f2}) automatically ensures the universality of 
gauge charges is also important.

In summary, the particle contents appearing in the low-energy 
effective theory on the domain wall are identical to those in 
the SM. All the KK modes can be sufficiently separated from the 
SM particles as long as we set $\Omega \sim 10^{2+a}$ GeV be 
sufficiently large. 
Nevertheless, our model is distinguishable from the SM by the 
new tree-level decay $h\to\gamma\gamma$ ($h\to gg$) 
and a finite electroweak monopole.
A possible concern in our model is the additional massless 
particle $Y(x)$ which is inevitable because it is the NG mode
for spontaneously broken translational symmetry.
However, thanks to the low-energy theorems, all the interactions 
including $Y(x)$ must appear with derivatives $\p_\mu Y(x)$.
Consequently, they are suppressed by the large mass 
scale $\Omega$ and have practically no impact on phenomena 
at energies much lower than the large mass scale $\Omega$. 
The KK quark pair production via NG particle exchange 
gives a lower bound for $\Omega$ which is larger than $1.5$~TeV. 
Larger $\Omega$ requires severer fine-tuning, but is safer 
phenomenologically, whereas smaller $\Omega$ requires less 
fine-tuning and can be disproved more easily by experimental 
data. 

Let us discuss possible effects of radiative corrections in 
our low-energy effective theory. 
The particle content of effective theory below the mass 
scale $\Omega$ is identical to SM except the NG boson $Y(x)$ 
for translation. 
The higher dimensional operators of NG boson interactions 
are suppressed by powers of the large mass $\Omega$. 
Hence they do not contribute for phenomena at energies much below 
the scale $\Omega$, in the spirit of effective Lagrangian approach. 
The only possible exception is the Higgs coupling of gauge 
fields expressed by higher dimensional operators with the small 
mass scale $\mu$ in the gauge kinetic function. 
This coupling of Higgs boson and gauge fields such as in 
Eq.~(\ref{eq:higgs_gauge_int}) is given by Higgs vacuum expectation 
value $v_h$. 
We need to assume that the higher dimensional coupling 
of Higgs boson and gauge fields are fine-tuned to that value 
when the Higgs vacuum expectation value is fine-tuned to a value 
much smaller than $\Omega$. 
With this assumption, we expect that the radiative corrections 
to quantities such as physical Higgs boson mass should be 
essentially the same as nonsupersymmetric SM. 
For instance we need to implement supersymmetry if we wish 
to make the fine tuning less severe in our model.

Models with warped spacetime \cite{Randall:1999ee,Randall:1999vf} 
 exhibit features similar to our model, except that the usual 
assumption of delta-function-like brane in models with warped 
spacetime is replaced by a smooth localized energy density 
(fat brane) in our model. 
Previously we have studied BPS domain-wall solutions embedded 
into four- and five-dimensional supergravity \cite{Eto:2002ns,
Eto:2003ut,Eto:2003bn}. 
These solutions are quite similar to BPS domain wall solutions in 
our present model. 
From these examples, we expect that our model can be coupled to 
gravity giving a fat brane embedded into warped spacetime. 
The resulting model should give physics in warped spacetime 
with finite wall width. 
We expect that phenomenology of our model will not be affected 
too much as long as we consider phenomena at energies 
below the gravitational (Planck) scale.

Finally, our model offers an interesting problem for 
the study of the cosmological evolution of the universe. 
Let us restrict ourselves in the region of temperature around 
the scale $\lambda\bar v \sim 10^2$GeV, where the analysis using effective 
potential is applicable. 
As we calculate explicitly in Appendix \ref{app:mode_eq} and 
\ref{sc:modeEq_unstable_sol}, we find that the 
effective potential computed on the stable background with 
$\langle H\rangle =v_h/\sqrt{2}$ is slightly different from 
that computed on the 
unstable background with vanishing Higgs $\langle H\rangle=0$. 
More explicitly, only the quadratic term has different 
coefficient $\lambda_2^2$ : it changes from $-4(\lambda\bar v)^2/3$ 
at $\langle H\rangle =v_h/\sqrt{2}$ to $-(\lambda\bar v)^2$. 
We can understand this phenomenon as follows. 
The definition of effective Higgs field depends on the 
background solution on which we expand the quantum fluctuation. 
The off-shell extrapolation of the effective potential 
computed on a particular background is different from that computed 
on a different background. 
Consequently, even though the extrapolated effective potential 
can give the position of another neighboring stationary 
point correctly, the curvature (mass squared) around it 
need not reproduce the value of mass squared computed on 
that point, since the background is different. 
This feature is in contrast to ordinary local field theory, 
and perhaps can be interpreted as a composite nature of 
fluctuation fields on solitons. 
The coefficient $\lambda_2^2$ is directly related to 
the transition temperature of phase transition during the 
cosmological evolution. 
At zero temperature, our effective potential 
(\ref{eq:constant_higgs}) calculated on the background of 
$\langle H\rangle =v_h/\sqrt{2}$ is valid, 
since we assume $\Omega<\lambda v$. 
As we heat up the universe starting from this situation, 
finite temperature effects come in to raise the effective 
potential for nonzero values of Higgs field. 
Eventually around a certain temperature of order $\lambda\bar v$, 
we will find a phase transition to the phase without 
Higgs condensation, namely $SU(2)\times U(1)$ gauge symmetry 
restoration. 
To estimate this transition temperature, we need to study the 
change of effective potential during this process. 
As we noted, the coefficient $\lambda_2^2$ is likely to change 
gradually from $-4(\lambda\bar v)^2/3$ to $-(\lambda\bar v)^2$. 
Therefore we need to take account of the change of $\lambda_2^2$ 
besides the finite temperature effects. 
This is an interesting new challenge to determine the transition 
temperature in this kind of models. 
We leave this issue for a future study.


\acknowledgements

This work is supported in part by the Albert Einstein Centre for Gravitation and Astrophysics financed by the Czech Science Agency Grant No. 14-37086G (F.\ B.).
This work is also supported in part 
by the Ministry of Education,
Culture, Sports, Science (MEXT)-Supported Program for the Strategic
Research Foundation at Private Universities ``Topological Science''
(Grant No.~S1511006) (N.\ S.), 
by the Japan Society for the 
Promotion of Science (JSPS) 
Grant-in-Aid for Scientific Research
(KAKENHI) Grant Numbers   
26800119, 16H03984 and 17H06462 (M.\ E.), and 
by the program of Czech Ministry of Education 
Youth and Sports INTEREXCELLENCE Grant number LTT17018 (F. B.).
F.\ B.\ was an international research fellow of the Japan Society 
for the Promotion of Science, and was supported by Grant-in-Aid 
for JSPS Fellows, Grant Number 26004750.

\appendix

\section{Mode equations on the stable BPS solution 
} 
\label{app:mode_eq}

Here we define mode expansions for Higgs and other fields 
in order to compute low-energy effective action in four dimensions. 
We need to choose a solution of field equations as a 
background on which we expand fluctuation fields. 
Since we are interested in the parameter region 
(\ref{eq:parameter_region}), we should choose the stable 
BPS solution in Eq.(\ref{eq:T1}). 
With this background, we define fluctuation fields $\delta T$ 
and $\delta H_R, \delta H_I$ as 
\begin{equation}
{\cal T}= v\tanh \Omega y+\frac{\delta {\cal T}}{\sqrt{2}}, 
\quad 
{\cal H}=\frac{\bar v}{\cosh\Omega y}
+\frac{\delta{\cal H}_R+i\delta{\cal H}_I}{\sqrt2}. 
\label{eq:fluctuation_unst}
\end{equation}
The quadratic part of the bosonic Lagrangian is given by means of 
Hamiltonians $K_{TR}, K_I$ 
\begin{equation}
{\cal L}^{(2)}=
{\cal L}^{(2)}_{TR}
+
{\cal L}^{(2)}_I,
\label{eq:quadratic_Lag}
\end{equation}
\begin{eqnarray}
{\cal L}^{(2)}_{TR}&=&\frac{1}{2}\Phi^T
[-\partial_\mu\partial^\mu-K_{T,R}]\Phi, 
\quad \Phi^T
=
(\delta {\cal T}, \delta{\cal H}_R), 
\nonumber \\
K_{TR}&=&-\partial_y^2{\bf 1}_2+
\left(
\begin{array}{cc}
2\lambda^2({\cal H}_0^2+3{\cal T}_0^2-v^2) & 4\lambda^2{\cal T}_0{\cal H}_0 \\
4\lambda^2{\cal T}_0{\cal H}_0 & \Omega^2+2\lambda^2(
3{\cal H}_0^2+{\cal T}_0^2-v^2)
\end{array}
\right),
\label{eq:hamitonian_TR}
\end{eqnarray}
\begin{eqnarray}
{\cal L}^{(2)}_I&=&\frac{1}{2}\delta {\cal H}_I
[-\partial_\mu\partial^\mu-K_I]\delta {\cal H}_I, 
\nonumber \\
K_I&=&-\partial_y^2+\Omega^2+2\lambda^2({\cal H}_0^2+{\cal T}_0^2-v^2).
\label{eq:hamitonian_HI}
\end{eqnarray}
Once we obtain eigenfunctions of these Hamiltonians, we can 
obtain mode expansions of the 5D fields into KK towers of 
effective fields, such as 
\begin{equation}
\Phi_i(x,y)=\sum_{n=0}^\infty \phi_n(x) u_i^{(n)}(y), 
\quad i=T, R. 
\label{mode_exp}
\end{equation}
where the $n$-th eigenstate generally has components in both 
5D fields $\delta{\cal T}$ and $\delta{\cal H}_R$, since 
they have coupled Hamiltonian $K_{TR}$. 
The label of eigenstates $n$ contains also continuum states. 

Since the $\delta{\cal H}_I$ will be absorbed by the gauge boson 
by the Higgs mechanism, we will consider only the coupled 
linearized field equation for $\delta {\cal T}$ and 
$\delta{\cal H}_R$. 
Since the coupled equation is difficult to solve exactly, we 
solve it starting from the $\lambda \bar v=0$ case as a perturbation 
series in powers of the small parameter 
$\epsilon^2=(\lambda\bar v/\Omega)^2$. 

At $\lambda \bar v=0$, 
the Hamiltonian $K_{TR}$ becomes diagonal and 
the ${\cal T}$ and ${\cal H}_R$ linearized field equations 
decouple 
\begin{equation}
K_T=-\partial_y^2+4\Omega^2-\frac{6\Omega^2}{\cosh^2\Omega y}, 
\label{eq:hamiltonian_T_unst}
\end{equation}
\begin{equation}
K_R=-\partial_y^2+\Omega^2-\frac{2\Omega^2}{\cosh^2\Omega y}. 
\label{eq:hamiltonian_H_unst}
\end{equation}
Eigenvalues of the Hamiltonian give mass squared $m^2$ of 
the corresponding effective fields. 

In the parameter region (\ref{eq:parameter_region}), 
we find two discrete bound states for $\delta T$, and a 
continuum of states with the threshold at 
$(m_T^{(2)})^2=(2\Omega)^2$ 
\begin{equation}
u_T^{(0)}(y)= \frac{\sqrt{3\Omega}}{2} 
\frac{1}{\cosh^2\Omega y}, 
\quad 
(m_T^{(0)})^2=0, 
\label{NG_mode}
\end{equation}
\begin{equation}
u_T^{(1)}(y)= \sqrt{\frac{3\Omega}{2}} 
\frac{\tanh\Omega y}{\cosh\Omega y}, 
\quad 
(m_T^{(1)})^2=3(\Omega)^2.
\label{T1_mode}
\end{equation}
We recognize that the massless mode is precisely the 
Nambu-Goldstone boson for spontaneously broken translation. 
For the fluctuation $\delta {\cal H}_R$, we find that there is 
only one discrete bound state below the threshold at $\Omega^2$ 
\begin{equation}
u_R^{(0)}(y)= \frac{\sqrt{\Omega}}{2} 
\frac{1}{\cosh\Omega y}, 
\quad 
(m_R^{(0)})^2=0.
\label{condensation_mode}
\end{equation}
This is the massless particle at the critical point where 
condensation of ${\cal H}_R$ starts. 
It is not an accident that the functional form of this mode 
function is identical to the condensation of ${\cal H}_R$ in 
Eq.~(\ref{eq:T1}). 
This mode will become massive physical Higgs particle when we 
switch on the perturbation $(\lambda\bar v)^2 >0$.  

We can now systematically compute the perturbative corrections 
in powers of small parameter $\epsilon$. 
The lowest order correction to the eigenvalue can be obtained 
by taking the expectation value of the perturbation Hamiltonian 
in terms of the lowest order wave function. 
Therefore we obtain the mass eigenvalue of the physical Higgs 
particle up to the leading order 
\begin{equation}
(m_h)^2=\int dy u_R^{(0)}(y)[(K_{TR})_{22}-K_R]u_R^{(0)}(y)
=\frac{8}{3}(\lambda \bar v)^2. 
\label{eq:mass2_higgs}
\end{equation}
This result agrees with the result of the analysis using the 
effective potential (\ref{eq:L_higgs}). 
In fact, we can reproduce the effective potential by evaluating 
the cubic and quartic terms in fluctuation field $\delta {\cal H}_R$. 
With the perturbation theory, we can compute corrections 
to the Higgs mass to any desired order of $\epsilon$. 
\begin{equation}
V_H =  -\frac{4(\lambda \bar v)^2}{3} |H|^2 + \frac{\lambda^2\Omega}{3} |H|^4,
\label{eq:eff_pot_stable}
\end{equation}
in agreement with Eq.~(\ref{eq:L_higgs}).

\section{Mode equations on the unstable BPS solution}
\label{sc:modeEq_unstable_sol}

We can choose another BPS solution (\ref{eq:T0}) as background, 
which becomes stable in the parameter region $\Omega< \lambda v$. 
We define a small fluctuation around this background as 
\begin{equation}
{\cal T}= v\tanh\lambda v y+\delta {\cal T}'/\sqrt{2}, 
\quad 
{\cal H}=(\delta{\cal H}'_R+i\delta{\cal H}'_I)/\sqrt2. 
\label{eq:fluctuation_unst}
\end{equation}
The linearized field equation, in this case, is decoupled 
with the Hamiltonian $K'_T,K'_R, K'_I$ as  
\begin{equation}
K'_T=-\partial_y^2+4(\lambda v)^2-\frac{6(\lambda v)^2}{\cosh^2\lambda v y}, 
\label{eq:hamiltonian_T_unst}
\end{equation}
\begin{equation}
K'_R=K'_I=-\partial_y^2+\Omega^2-\frac{2(\lambda v)^2}{\cosh^2\lambda v y}.
\label{eq:hamiltonian_H_unst}
\end{equation}
We find exact mode functions in this case. 
We find two discrete bound states for $\delta {\cal T}'$ 
and a continuum of states with the threshold at 
$(m'{}_T^{(2)})^2=(2\lambda v)^2$ 
\begin{equation}
u'{}_T^{(0)}(y)= \frac{\sqrt{3\lambda v}}{2} 
\frac{1}{\cosh^2\lambda v y}, 
\quad 
(m'{}_T^{(0)})^2=0, 
\label{NG_mode_unst}
\end{equation}
\begin{equation}
u'{}_T^{(1)}(y)= \sqrt{\frac{3\lambda v}{2}} 
\frac{\tanh\Omega y}{\cosh\lambda v y}, 
\quad 
(m'{}_T^{(1)})^2=3(\lambda v)^2.
\label{T1_mode_unst}
\end{equation}
The massless mode gives an exact NG boson mode function in this case. 
For the fluctuation $\delta {\cal H}'_R$, we find that there is 
only one discrete bound state below the threshold at $\Omega^2$ 
\begin{equation}
u'{}_R^{(0)}(y)= \frac{\sqrt{\lambda v}}{2} 
\frac{1}{\cosh\lambda v y}, 
\quad 
(m'{}_R^{(0)})^2=-(\lambda \bar v)^2. 
\label{condensation_mode}
\end{equation}
This is precisely the tachyonic mode at the unstable background 
solution. 
We note that the value of (negative) mass squared is different 
from the corresponding value $-4(\lambda\bar v)^2/3$ 
of the off-shell extension to $H=0$ 
of the effective potential computed on the stable BPS solution in 
Eq.(\ref{eq:constant_higgs}). 
This is due to the fact that a different background solution 
gives a different spectrum of fluctuations, even though they are 
qualitatively similar.

Once the exact mode function is obtained, on the background of 
the unstable solution, we only need to insert the following 
Ansatz into the 5D Lagrangian and integrate over $y$, in order 
to obtain the effective potential of the effective Higgs field 
$H'(x)$. 
\begin{equation}
{\cal T}= v\tanh\lambda v y, \quad 
{\cal H}=H'(x)\sqrt{\frac{\lambda v}{2}}\frac{1}{\cosh \lambda v y}. 
\label{eq:ansatz_unstable}
\end{equation}
After integrating over $y$, we obtain the effective action as 
\begin{equation}
{\cal L}_{\rm Higgs}(H')
= |D_\mu H'|^2  
- V_{H'}, 
\quad
V_{H'} = - (\lambda \bar v)^2 |H'|^2 + \frac{\lambda^2\Omega}{3} |H'|^4.
\label{eq:L_higgs_unst}
\end{equation}
The quadratic term agrees with the mass squared eigenvalue of 
the mode equation of fluctuations. 
It is interesting to observe that the coefficient of the quadratic 
term is different from that computed on the stable BPS solution as 
background,  although the quartic term is identical. 

\section{Cross section for KK fermion pair production by NG boson exchange}
\label{sc:cross}

Here we calculate the differential cross section (\ref{eq:cross1}). 
First we consider the process $u_Ld_L\rightarrow u_L^{(1)}d_{L}^{(1)}$, 
whose Feynman diagram is shown in Fig. \ref{fig:diagram}(a). 
The amplitude is given in terms of spinor wave functions $u_{uL}$ 
and $u_{dL}$ of incoming SM fermions, and $u_{u^{(1)}L}$ and 
$u_{d^{(1)}L}$ of outgoing KK quarks as 
\begin{eqnarray}
 i{\cal M}={\alpha^2 \Omega \over v^2}{i \over t}
(\bar{u}_{u^{(1)}L}(k_1)i(\slashed{p}_1-\slashed{k}_1)u_{uL}(p_1))
(\bar{u}_{d^{(1)}L}(k_2)i(\slashed{p}_1-\slashed{k}_1)u_{dL}(p_2)),
\end{eqnarray}
with $t=(p_1-k_1)^2$. 
We approximate SM quarks to be massless, and assume the same 
vertex couplings $\alpha$ for $uu^{(1)}Y$ and $dd^{(1)}Y$ for 
simplicity, although they can be different since fermion wave 
functions for $u, u^{(1)}$ and $d, d^{(1)}$ are in general 
different. 
The squared amplitude is
\begin{eqnarray}
 |{\cal M}|^2&=&{4\alpha^4 \Omega^2 \over v^4 t^2}
\left\{2(p_1\cdot (p_1-k_1))(k_1\cdot (p_1-k_1))-(p_1\cdot k_1)t 
\right\} \nonumber \\
 &&
 \quad \quad \times \left\{2(p_2\cdot (p_1-k_1))
(k_2\cdot (p_1-k_1))-(p_2\cdot k_2)t  \right\}  \nonumber \\
 &=& {4\alpha^4 \Omega^2 \over v^4}
{E^4(1-\beta \cos\theta)^2(1-\beta^2) \over (\beta^2+1-2\beta\cos\theta)^2},
\end{eqnarray}
which leads to the differential cross section 
\begin{eqnarray}
{d\sigma \over dt}(u_Ld_L\rightarrow u_L^{(1)}d_{L}^{(1)})
={\alpha^4 \over 576 \pi^2 s}{\Omega^2 \over v^4}
{E^2(1-\beta_{M_1}\cos\theta)^2(1-\beta_{M_1}^2) \over 
(\beta_{M_1}^2+1-2\beta_{M_1}\cos\theta)^2}, \label{eq:cross3}
\end{eqnarray}
with $s = 4 E^2$.
Other combinations of initial quark chiralities $RR, LR, RL$ 
are found to give identical differential cross sections. 
Hence we find (\ref{eq:cross1}).


\end{document}